\documentclass[aps,pra,reprint,amsmath,amsfonts,amssymb,superscriptaddress]{revtex4-1}
\usepackage{graphicx,float,calc}
\usepackage{color,bm}
\usepackage{ulem}
\usepackage{braket}
\usepackage[colorlinks,urlcolor=blue,citecolor=blue,linkcolor=blue]{hyperref}
\begin{document}

\title{Quantum criticality and Kibble-Zurek scaling in the Aubry-Andr\'{e}-Stark model}

\author{En-Wen Liang}
\affiliation{Key Laboratory of Atomic and Subatomic Structure and Quantum Control (Ministry of Education), Guangdong Basic Research Center of Excellence for Structure and Fundamental Interactions of Matter, South China Normal University, Guangzhou 510006, China}
\affiliation{Guangdong Provincial Key Laboratory of Quantum Engineering and Quantum Materials,School of Physics, South China Normal University, Guangzhou 510006, China}

\author{Ling-Zhi Tang}
\affiliation{Key Laboratory of Atomic and Subatomic Structure and Quantum Control (Ministry of Education), Guangdong Basic Research Center of Excellence for Structure and Fundamental Interactions of Matter, South China Normal University, Guangzhou 510006, China}
\affiliation{Guangdong Provincial Key Laboratory of Quantum Engineering and Quantum Materials,School of Physics, South China Normal University, Guangzhou 510006, China}

\author{Dan-Wei Zhang}
\email{danweizhang@m.scnu.edu.cn}
\affiliation{Key Laboratory of Atomic and Subatomic Structure and Quantum Control (Ministry of Education), Guangdong Basic Research Center of Excellence for Structure and Fundamental Interactions of Matter, South China Normal University, Guangzhou 510006, China}
\affiliation{Guangdong Provincial Key Laboratory of Quantum Engineering and Quantum Materials,School of Physics, South China Normal University, Guangzhou 510006, China}

\begin{abstract}
We explore quantum criticality and Kibble-Zurek scaling (KZS) in the Aubry-Andr\'{e}-Stark (AAS) model, where the Stark field of strength $\varepsilon$ is added onto the one-dimensional quasiperiodic lattice. We perform scaling analysis and numerical calculations of the localization length, inverse participation ratio (IPR), and energy gap between the ground and first excited states to characterize critical properties of the delocalization-localization transition. Remarkably, our scaling analysis shows that, near the critical point, the localization length $\xi$ scales with $\varepsilon$ as $\xi\propto\varepsilon^{-\nu}$ with $\nu\approx0.3$ a new critical exponent for the AAS model, which is different from the counterparts for both the pure Aubry-Andr\'{e} (AA) model and Stark model. The IPR $\mathcal{I}$ scales as $\mathcal{I} \propto \varepsilon^{s}$ with the critical exponent $s\approx0.098$, which is also different from both two pure models. The energy gap $\Delta E$ scales as $\Delta E\propto \varepsilon^{\nu z}$ with the same critical exponent $z\approx2.374$ as that for the pure AA model. We further reveal hybrid scaling functions in the overlap between the critical regions of the Anderson and Stark localizations. Moreover, we investigate the driven dynamics of the localization transitions in the AAS model. By linearly changing the Stark (quasiperiodic) potential, we calculate the evolution of the localization length and the IPR, and study their dependence on the driving rate. We find that the driven dynamics from the ground state is well described by the KZS with the critical exponents obtained from the static scaling analysis. When both the Stark and quasiperiodic potentials are relevant, the KZS form includes the two scaling variables. This work extends our understanding of critical phenomena on localization transitions and generalizes the application of the KZS to hybrid models.

\end{abstract}

\date{\today}

\maketitle

\section{Introduction}

In recent years, there has been increasing interest in the studies of Anderson localization \cite{Anderson1958,Abrahams1979,Patrick1985} and localization transitions in quasiperiodic systems \cite{Harper1955,Aubry1980,lellouch2014localization,devakul2017anderson,roy2022critical,agrawal_universality_2020,goblot_emergence_2020,roy_critical_2022,agrawal_quasiperiodic_2022}. Compared to random systems with quenched disorders, quasiperiodic systems exhibit unique localization properties that have been explored both theoretically \cite{goblot_emergence_2020,agrawal_universality_2020,roy_critical_2022,agrawal_quasiperiodic_2022} and experimentally \cite{crespi_anderson_2013,bordia_coupling_2016,roati_anderson_2008}. The one-dimensional Aubry-Andr\'{e} (AA) model \cite{Aubry1980} serves as an important example in this regard, where a localization transition occurs when the strength of the quasiperiodic potential exceeds the critical point determined by the self-duality \cite{Biddle2010,liu_generalized_2020,wang_engineering_2023}. Furthermore, various extensions of the AA model have been proposed to investigate the mobility edges \cite{Soukoulis1982,Ganeshan2015}, topological phases \cite{DWzhang2018,DWZhang2020,yoshii_topological_2021,GQZhang2021,tang_topological_2022,wang_topological_2022,nakajima_competition_2021,wu_quantized_2022,SHuang2024}, many-body localization \cite{noauthor_observation_nodate,lukin_probing_2019,wang_non-hermitian_2023}, and critical phenomena  \cite{goblot2020emergence,lv_exploring_2022,lv_quantum_2022,aramthottil_finite-size_2021,PhysRevB.96.214201,S.Yin2022a, S.Yin2022b,Roosz2024}. Remarkably, the quantum criticality and scaling functions with new critical exponents for the localization transition in the disordered AA model have been unearthed in Ref. \cite{S.Yin2022a}, where the random disorder contributes an independent relevant direction near the AA critical point. According to the renormalization-group theory \cite{belitz_how_2005,fisher_renormalization_1974,gosselin_renormalization_2001,zhong_probing_2006}, critical exponents in the scaling functions of physical observables around the critical point \cite{cherroret_how_2014,lemarie_universality_2009,slevin_critical_2009,you_fidelity_2011,su_role_2018,wang_berezinskii-kosterlitz-thouless-like_2024,slevin_anderson_1997,asada_anderson_2002,rams_quantum_2011} characterize the universal features of continuous quantum (classical) phase transitions \cite{Sachdev2011,Osterloh2002,Heyl2018,carollo_geometry_2020,vojta2003quantum}. Thus, determining critical exponents is crucial in understanding critical phenomena and phase transitions, including the localization transition.

On the other hand, the Kibble-Zurek mechanism \cite{kibble1976topology,zurek1985cosmological,kibble_implications_1980,zurek_cosmological_1996} provides a powerful framework to investigate critical dynamics of phase transitions ranging from cosmology to condensed matter systems \cite{Laguna1997,Zurek2005,Polkovnikov2005,dziarmaga2010dynamics,Polkovnikov2011,Campo2014,Das2012,Bando2020,HBZeng2023,qiu_observation_2020,anquez_quantum_2016,MGong2016,ZPGao2017,PhysRevB.90.134108,qiu2020observation}.
Based on this framework, driven dynamics across a critical point can be described by the universal Kibble-Zurek scaling (KZS) and the critical exponents associated with the phase transition can be extracted. Recently, more and more attention has been paid to the dynamics in disorder-driven  transitions \cite{CWLiu2015,Roosz2014,Morales-Molina2014,Serbyn2014,Sinha2019,PhysRevB.106.134203,PhysRevB.103.104202,Decker2020,S.Yin2022b,S.Yin2022c}. In particular, the KZS has been generalized to characterize the driven dynamics in the disordered AA model \cite{S.Yin2022a}, which can include two scaling variables when both the random and quasiperiodic potentials are relevant directions \cite{S.Yin2022b}. Notably, random and quasiperiodic disorders are not the only route to induce localization transitions. For instance, localization can also manifest in systems featuring a linear potential, known as the Wannier-Stark localization in the noninteracting case \cite{Wannier1960,Emin1987,zhang_mobility_2021,taylor_experimental_2020,lang_disorder-free_2022,schmidt_signatures_2018,guo_observation_2021}. In the presence of interactions, the Stark many-body localization has been revealed \cite{Schulz2019,van_nieuwenburg_bloch_2019,guo_stark_2021,wang_stark_2021,morong_observation_2021,wei_static_2022,van_nieuwenburg_bloch_2019}.
It has been shown that the Stark field can induce to diffusive dynamics under the interplay between Anderson and Stark localizations in two-dimensional random lattices \cite{kolovsky_interplay_2008}. Moreover, the weak-field sensing with super-Heisenberg precision based on the Stark localization has been proposed \cite{PhysRevLett.131.010801}. However, the critical properties and related KZS near localization transitions in the presence of the Stark and quasiperiodic fields remain unexplored.

In this article, we investigate the quantum criticality and the KZS in the Aubry-Andr\'{e}-Stark (AAS) model, where the Stark field is imposed onto the one-dimensional quasiperiodic lattice. We perform scaling analysis and numerical calculations of the localization length, inverse participation ratio (IPR), and energy gap between the ground and first excited states to reveal exotic critical properties of the localization transition. Remarkably, our scaling analysis of the localization length and the IPR near the critical point shows two new critical exponents for the AAS model, which are different from the counterparts for both the pure AA model and Stark model. In contrast, the scaling form of the energy gap shares the same critical exponent as that for the pure AA model. We further obtain hybrid scaling functions in the overlap between the critical regions of the Anderson and Stark localizations. Moreover, we explore the driven dynamics of the localization transitions in the pure AA and Stark models and the AAS model. By linearly quenching the strength of the Stark or quasiperiodic potential, we calculate the evolution of the localization length and the IPR under various driving rates. We find that the driven dynamics from the ground state is well described by the KZS with the critical exponents obtained from the static scaling analysis. When both the Stark and quasiperiodic potentials are relevant directions, the KZS form contains the two scaling variables.

The rest of the paper is organized as follows. In Sec.~\ref{sec2}, we introduce the AAS model and the method of the scaling analysis. In Sec.~\ref{sec3}, we investigate the critical properties of localization transitions for the pure AA model, Stark model, and the AAS model, respectively. The scaling forms with new exponents for the AAS model are obtained. Sec.~\ref{sec4} is denoted to study the driven dynamics of the localization transitions by using the KZS. Finally in Sec.~\ref{sec5}, a brief conclusion is presented.

\section{\label{sec2} model and method}

We consider the AA model with a linear gradient field across the lattice of $L$ sites, which is described by the following AAS Hamiltonian:
\begin{equation}
\label{H}
\begin{aligned}
H_{\text{AAS}} = & -J\sum_{j}^{L-1}{(c_{j}^\dagger c_{j+1}+h.c.)} +\varepsilon \sum_{j}^{L-1}j c_j^\dagger c_j\\
& +(2J+\delta)\sum_{j}^{L-1}\cos{[2\pi(\gamma j+\phi)]c_j^\dagger c_j}.
\end{aligned}
\end{equation}
Here $c_j^\dagger$ $(c_j)$ represents the creation (annihilation) operator at site $j$, $J$ is the hopping strength, $\varepsilon$ and $2J+\delta$ denote the strengths of the Stark field and the quasi-periodic lattice, respectively. The lattice phase $\phi$ is uniformly chosen from the interval $[0,1]$ for averaging over the pseudorandom potentials. In the following, we set $J=1$ as the energy unit, choose the inverse golden mean $\alpha = (\sqrt{5}-1)/2=\lim_{j\rightarrow\infty}F_j/F_{j+1}$ to approach an incommensurate lattice via two consecutive Fibonacci sequences $F_j$ and $F_{j+1}=L$, and adopt open boundary conditions in our numerical calculations with the exact diagonalization method.

\begin{figure}[tb]
	\centering
	\includegraphics[width=0.4\textwidth]{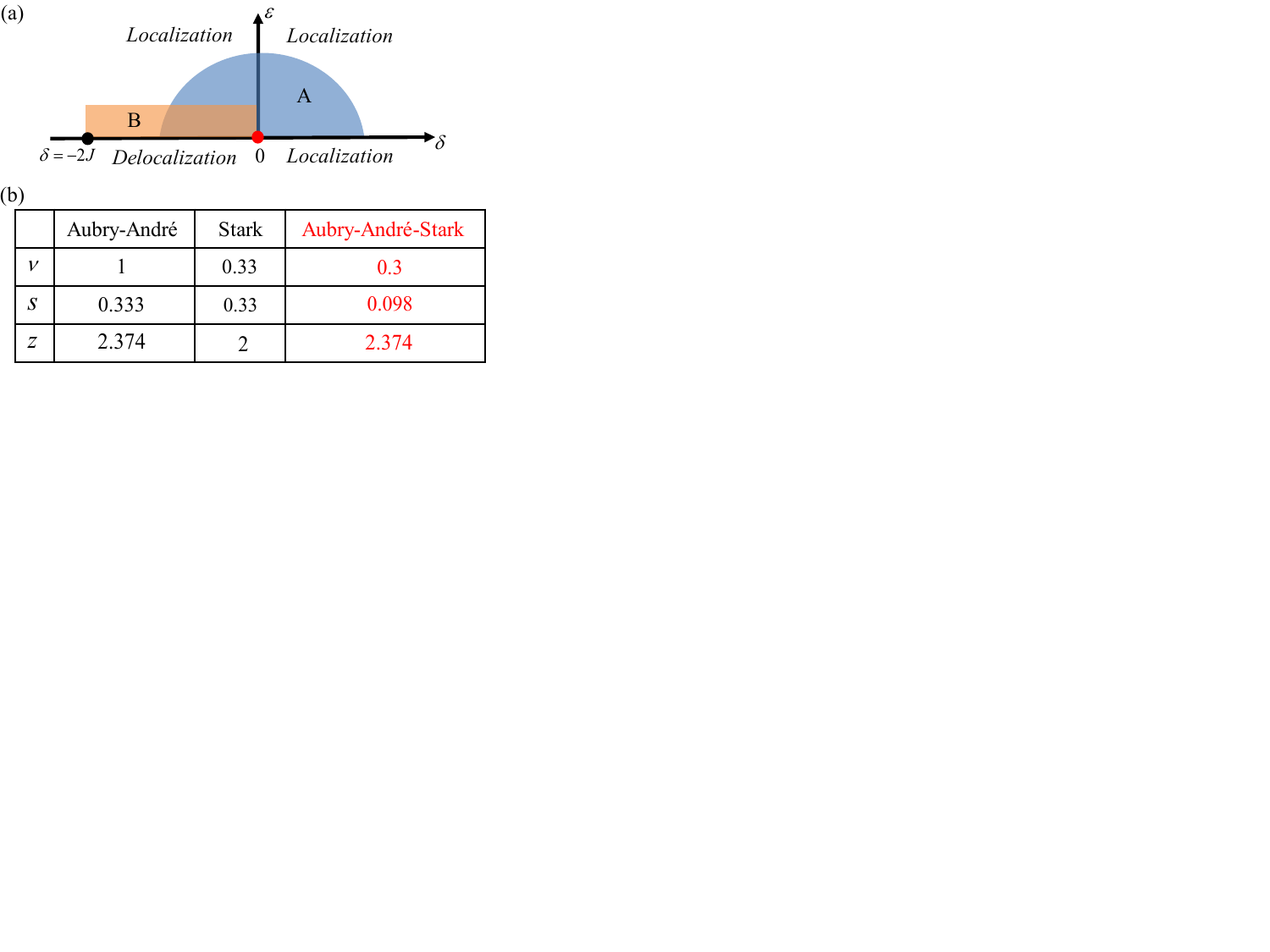}
	\caption{(Color online) (a) Sketch of the localization phase diagram of the AAS model. The blue region A denotes the critical region of localization transition of the AAS model. The orange region B denotes the critical region of the Stark localization transition. Near the critical point (denoted by the red point) at $\delta=\varepsilon=0$, the two critical regions overlap. When $\delta=-2J$ (denoted by the black point), the model returns to the pure Stark model. (b) The extracted critical exponents $\{\nu,s,z\}$ for the AA model, Stark model and AAS model.
	}\label{fig1}
\end{figure}

When $\varepsilon=0$, this model reduces to pure AA model with the critical point at $\delta=0$. All eigenstates are extended and localized for $\delta<0$ and $\delta>0$, respectively. When $\delta = -2J$, this model returns to pure Stark model. When $L\rightarrow\infty$, the Stark localization transition occurs at $\varepsilon = 0$ \cite{PhysRevLett.101.190602,van_nieuwenburg_bloch_2019,PhysRevLett.131.010801}, which means that all eigenstates will be localized under any finite Stark potential $\varepsilon$. Thus, we can sketch the localization phase diagram of the AAS model, as shown in Fig.~\ref{fig1}(a). Near the localization critical point $\delta=\varepsilon=0$, one has the critical region A for the AAS model with two variables $\delta$ and $\varepsilon$. In addition, for $\delta < 0$ and infinitesimal $\varepsilon$, there is a critical region B for the Stark localization. Since there is no mobility edge in the AAS model, we focus on the localization transition of the ground state and explore its quantum criticality and the KZS in the following sections.

At the critical point of the localization transition, the wave function of the ground state is neither localized nor extended. The occurrence of quantum phase transitions can be verified by several physical quantities. Here we use three characteristic physical quantities to explore the quantum criticality of the localization phase transition in the AAS model. The first quantity is the localization length $\xi$ given by
\begin{equation}
\label{Eq:xiscaling}
	\xi = \sqrt{\sum_{j>j_c}^{L} [( j - j_c )^2 ] |\psi(j)|^2},
\end{equation}
where $\psi(j)$ denotes the wave function of the ground state, and $j_c\equiv\sum j|\psi(j)|^2$ represents the localization center. Near a critical point of the delocalization-localization transition, $\xi$ scales with the distance to the critical point $g$ as
\begin{equation}
\label{Eq:xiscaling1}
\xi\propto g^{-\nu},
\end{equation}
where $\nu$ is the critical exponent. For the pure AA model, $g=\delta$ and $\nu=\nu_{\delta}=1$ ~\cite{PhysRevA.99.042117,Sinha2019,S.Yin2022c}. For the pure Stark model, $g=\varepsilon$ and $\nu=\nu_{\varepsilon}\approx0.33$ \cite{PhysRevLett.131.010801}.

The second quantity is the IPR defined as
\begin{equation}
\label{Eq:ipr}
{\mathcal{I} } = \frac{{\sum_{j=1}^L|\psi (j)|^4}}{{\sum_{j=1}^L|\psi (j)|^2}}.
\end{equation}
For the delocalization state, ${\mathcal{I}}$ scales as ${\mathcal{I}}\propto L^{-1}$ as the wave function is homogeneously distributed in the lattice. For the localization state, one has ${\mathcal{I}}\propto L^0$. At the critical point of the localization transition, ${\mathcal{I}}$ satisfies the following scaling relation with the lattice size $L$~
\begin{equation}
\label{Eq:iprscaling1}
{\mathcal{I} }\propto L^{-s/\nu}
\end{equation}	
with the critical exponent $s$. When $L\rightarrow\infty$, ${\mathcal{I}}$ scales with $g$ as
\begin{equation}
\label{Eq:iprscaling2}
{\mathcal{I}}\propto g^{s}.
\end{equation}

Finally, we consider the energy gap between the ground state and the first excited state $\Delta E$ to characterize the quantum criticality. At the localization transition point, the energy gap $\Delta E$ should scales as
\begin{equation}
\label{Eq:gapscaling1}
   \Delta E \propto L^{-z},
\end{equation}
according to the finite-size scaling with $z$ being the critical exponent. For the pure AA model, $z=z_{\delta}=2.374$ ~\cite{PhysRevA.99.042117,Sinha2019,S.Yin2022c}. For the pure Stark model, $z=z_{\varepsilon}=2$ \cite{PhysRevLett.131.010801}. When $L\rightarrow\infty$, $\Delta E$ scales with $g$ as
\begin{equation}
\label{Eq:gapscaling2}
{\Delta E}\propto g^{\nu z}.
\end{equation}

To explore the critical properties and numerically obtain the three critical exponents, we use the finite-size scaling, as summarized in Fig.~\ref{fig1}(b). The scaling analysis takes the ansatz
\begin{equation}
\label{Eq:generalscaling1}
   \ P(g) = L^{\rho/\nu} f(gL^{1/\nu}),
\end{equation}
where the physical quantities $P=\{\xi,\mathcal{I},\Delta E\}$, $f(.)$ is the scaling function, and $\rho$ denotes a critical exponent. When $L\rightarrow\infty$, it recovers the scaling relation
\begin{equation}
\label{Eq:generalscaling}
\ P(g) \propto g^{-\rho}
\end{equation}
for the three physical quantities.

\begin{figure}[tb]
\centering
\includegraphics[width=0.46\textwidth]{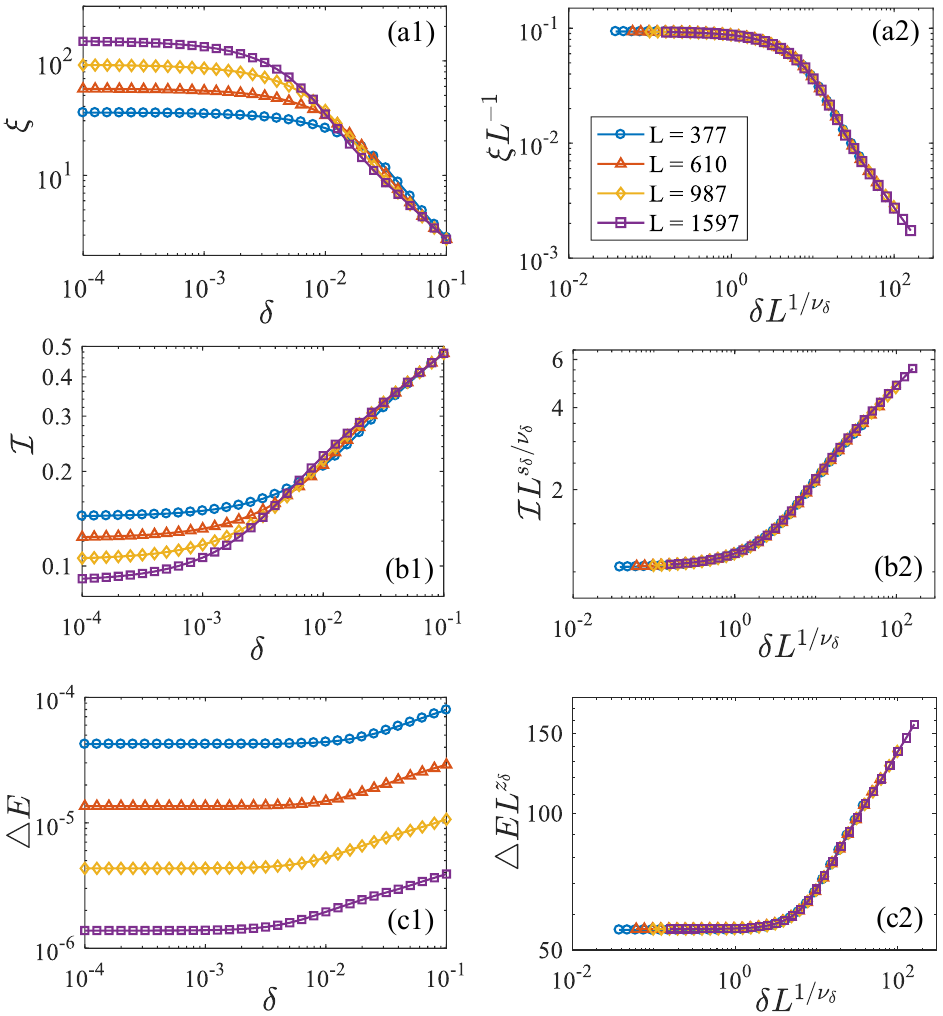}
\caption{Critical properties in the pure AA model with $\varepsilon=0$. Curves of $\xi$ versus $\delta$ before (a1) and after (a2) rescaling for different system sizes $L$. Curves of $\mathcal{I} $ versus $\delta$ before (b1) and after (b2) rescaling for various $L$. Curves of $\Delta E $ versus $\delta$ before (c1) and after (c2) rescaling for various $L$. The results are averaged over 1000 choices of $\phi$.}
\label{fig2}
\end{figure}

\section{\label{sec3}Quantum criticality}
In this section, we first perform scaling analysis in the pure AA model and Stark model, and obtain the corresponding critical exponents $\{\nu,s,z\}=\{\nu_{\delta},s_{\delta},z_{\delta}\}$ and $\{\nu_{\varepsilon},s_{\varepsilon},z_{\varepsilon}\}$, respectively. Then we investigate the critical properties in the AAS model and reveal new critical exponents.

\begin{figure}[tb]
	\centering
	\includegraphics[width=0.46\textwidth]{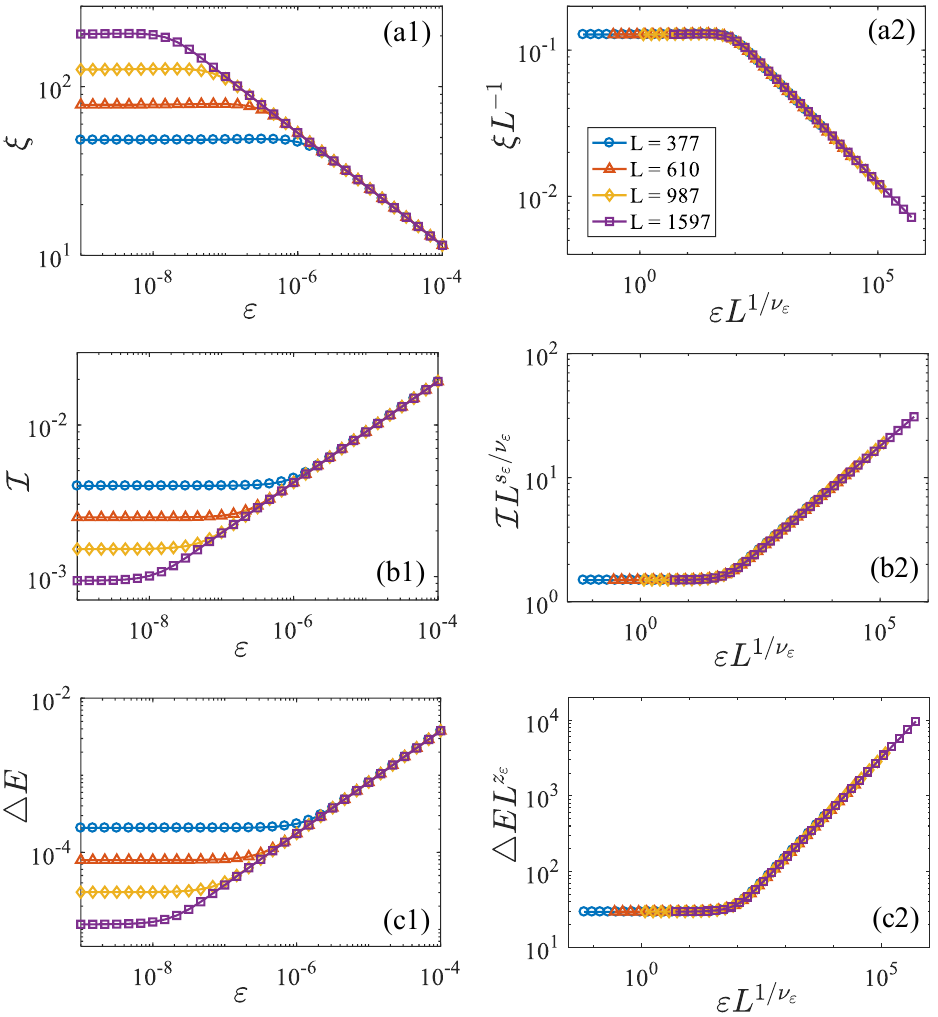}
	\caption{Critical properties in the pure Stark model with $\delta=-2J $. Curves of $\xi$ versus $\varepsilon$ before (a1) and after (a2) rescaling under for system sizes $L$. Curves of $\mathcal{I} $ versus $\varepsilon$ before (b1) and after (b2) rescaling for various $L$. Curves of $\Delta E $ versus $\varepsilon$ before (c1) and after (c2) rescaling for various $L$. The results are averaged over 1000 choices of $\phi$.}
	\label{fig3}
\end{figure}

\subsection{Pure \label{A}Aubry-Andr\'{e} and Stark criticalities}

When $\varepsilon=0$, our model returns to the pure AA model with the critical point at $\delta=0$. At this critical point, we can use the finite-size scaling to obtain the scaling functions for the three physical quantities $P=\{\xi,\mathcal{I},\Delta E\}$. For the localization length $\xi$, the scaling function can be derived from Eqs.~(\ref{Eq:xiscaling1},\ref{Eq:generalscaling1},\ref{Eq:generalscaling}) with $\rho =v_{\delta}$. The scaling analysis of $\xi$ should satisfy the following form
\begin{equation}
\label{Eq:xiscaling2}
\xi=Lf_1(\delta L^{1/\nu_\delta}),
\end{equation}
where $f_{i}(.)$ ($i=1$ here) is the scaling function. To determine the critical exponent $\nu_\delta$, we numerically calculate $\xi$ versus $\delta$ for various system size $L$ up to $L=1597$, as shown in Fig.~\ref{fig2}(a1). By rescaling $\xi$ and $\delta$ as $\xi L^{-1}$ and $\delta L^{1/\nu_{\delta}}$, we can estimate $\nu_{\delta}$ according to Eq.~(\ref{Eq:xiscaling2}). As shown in Fig.~\ref{fig2}(a2), the curves collapse onto each other very well when we choose $\nu_\delta=1$. Similarly, the finite-size scaling of the IPR $\mathcal{I}$ can be derived from Eqs.~(\ref{Eq:iprscaling2},\ref{Eq:generalscaling1},\ref{Eq:generalscaling}) with $\rho=-s_{\delta}$. The scaling form of $\mathcal{I}$ takes the form
\begin{equation}
    \label{Eq:iprscaling3}
\mathcal{I}=L^{-s_\delta/\nu_\delta} f_2(\delta L^{1/\nu_\delta}).
\end{equation}
The numerical results before and after rescaling $\mathcal{I}$ and $\delta$ as $\mathcal{I} L^{s_{\delta}/v_{\delta}}$ and $\delta L^{1/v_{\delta}}$ are shown in Figs.~\ref{fig2}(b1) and ~\ref{fig2}(b2), respectively. The best fitting for all the curves is obtained when $s_{\delta}=0.333$. By combing Eqs.~(\ref{Eq:gapscaling2},\ref{Eq:generalscaling1},\ref{Eq:generalscaling}) with $\rho =-z_{\delta}v_{\delta}$, we obtain the scaling function of the energy gap $\Delta E$ as
\begin{equation}
    \label{Eq:gapscaling3}
\Delta E=L^{-z_\delta} f_3(\delta L^{1/\nu_\delta}).
\end{equation}
The numerical results before and after rescaling $\Delta E$ and $\delta$ as $\Delta E L^{z_{\delta}}$ and $\delta L^{1/\nu_{\delta}}$ are shown in Figs.~\ref{fig2}(c1) and ~\ref{fig2}(c2), respectively. We find that the rescaled curves collapse one curve when $z_{\delta}=2.374$. The numerically obtained critical exponents $\{\nu,s,z\}=\{\nu_{\delta},s_{\delta},z_{\delta}\}$ for the pure AA model are summarized in Fig. \ref{fig1}(b), which are consistent with those in Refs. {\cite{PhysRevA.99.042117,Sinha2019,S.Yin2022c}}.

\begin{figure}[tb]
	\centering
	\includegraphics[width=0.46\textwidth]{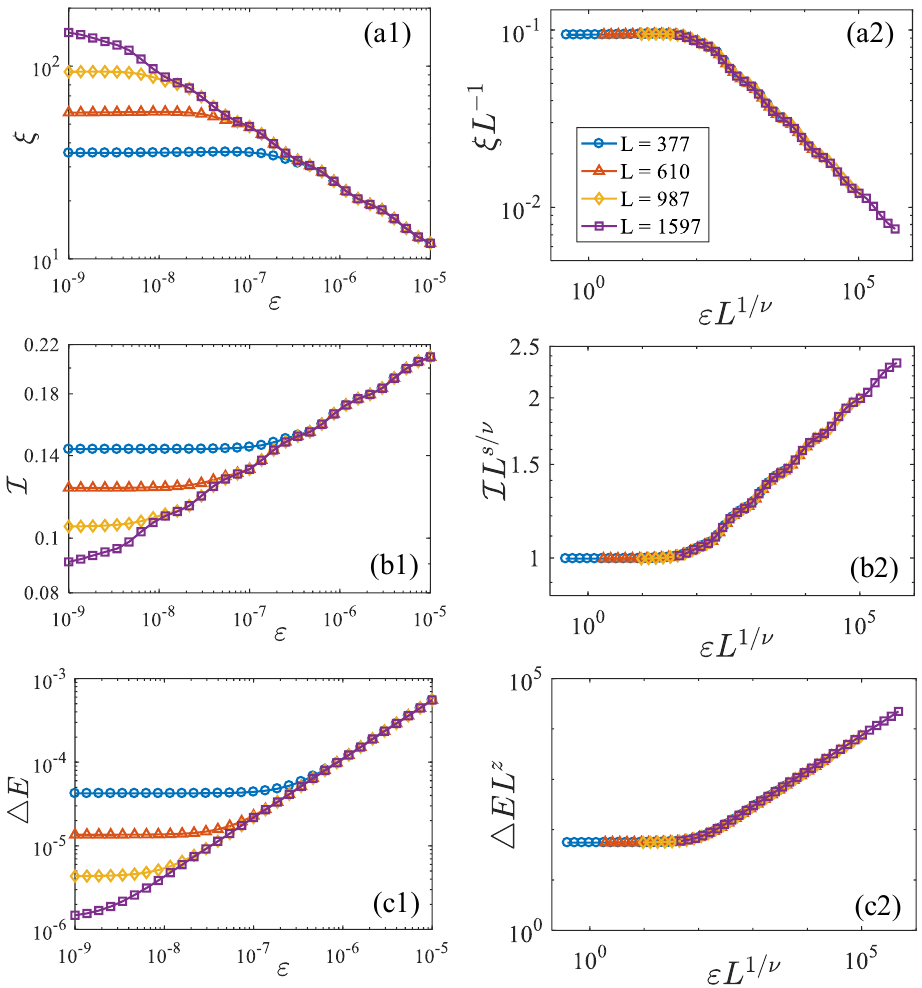}
	\caption{Critical properties in the AAS model at $\delta=0$. Curves of $\xi$ versus $\varepsilon$ before (a1) and after (a2) rescaling for different system sizes $L$. Curves of $\mathcal{I} $ versus $\varepsilon$ before (b1) and after (b2) rescaling for various $L$. Curves of $\Delta E $ versus $\varepsilon$ before (c1) and after (c2) rescaling for various $L$. The results are averaged over 1000 choices of $\phi$.}
	\label{fig4}
\end{figure}

When $\delta=-2J$, our model returns to the pure Stark model with the critical point at $\varepsilon=0$ when $L\rightarrow\infty$. Similarly, we perform the finite-size scaling of the three physical quantities $P=\{\xi,\mathcal{I},\Delta E\}$. For the localization length $\xi$, the scaling form can be derived as
\begin{equation}
\label{Eq:xiscaling3}
\xi=Lf_4(\varepsilon L^{1/\nu_\varepsilon}).
\end{equation}
The numerical results of $\xi$ versus the Stark field strength $\varepsilon$ for various $L$ are shown in Fig.~\ref{fig3}(a1). One can see that an initial flat region of $\varepsilon$ exists due to the delocalization nature of the wave function for finite $L$, which tends to be smaller (vanishing) as $\varepsilon$ is increased. The localization length becomes independent on the system size beyond a small threshold, which indicates the presence of the Stark localization. By rescaling $\xi$ and $\varepsilon$ as $\xi L^{-1}$ and $\varepsilon L^{1/\nu_{\varepsilon}}$ in Fig.~\ref{fig3}(a2), we find the best collapse for $\nu_{\varepsilon}=0.33$. Similarly, the IPR $\mathcal{I}$ in the pure Stark model satisfies the scaling form
\begin{equation}
\label{Eq:iprscaling4}
{\mathcal{I} }=L^{-s_\varepsilon/\nu_\varepsilon} f_5(\varepsilon L^{1/\nu_\varepsilon}).
\end{equation}
Figure \ref{fig3}(b1) shows the numerical results of $\mathcal{I}$ versus $\varepsilon$ for various $L$. The values of $\mathcal{I}$ in a region of small $\varepsilon$ are flat with approximate value of $L^{-1}$ for the delocalization state, while becomes independent of $L$ for the localization state beyond this region. To determine the critical exponent $s_{\varepsilon}$ in this case, we rescales $\mathcal{I}$ and $\varepsilon$ as $\mathcal{I} L^{s_{\varepsilon}/v_{\varepsilon}}$ and $\varepsilon L^{1/\nu_{\varepsilon}}$, and obtain the best collapse for $s_{\varepsilon} = 0.33$ in Fig.~\ref{fig3}(b2). The energy gap $\Delta E$ in the pure Stark model satisfies the scaling form
\begin{equation}
\label{Eq:gapscaling4}
\Delta E=L^{-z_\varepsilon} f_6(\varepsilon L^{1/\nu_\varepsilon}),
\end{equation}
Figure \ref{fig3}(c1) shows $\Delta E$ versus $\varepsilon$ for different lattice sizes. A flat region is also exhibited with a small energy gap for the delocalization state. In the localization region, $\Delta E$ becomes independent of $L$ and larger as $\varepsilon$ is increased. The critical exponent $z_\varepsilon$ is determined through the data collapse in Fig.~\ref{fig3}(c2), which yields $z_\varepsilon=2$. The numerically obtained critical exponents $\{\nu,s,z\}=\{\nu_{\varepsilon},s_{\varepsilon},z_{\varepsilon}\}$ for the pure Stark model are summarized in Fig. \ref{fig1}(b), which are consistent with those in Refs. {\cite{PhysRevLett.131.010801}
	
\begin{figure}[tb]
	\centering
	\includegraphics[width=0.46\textwidth]{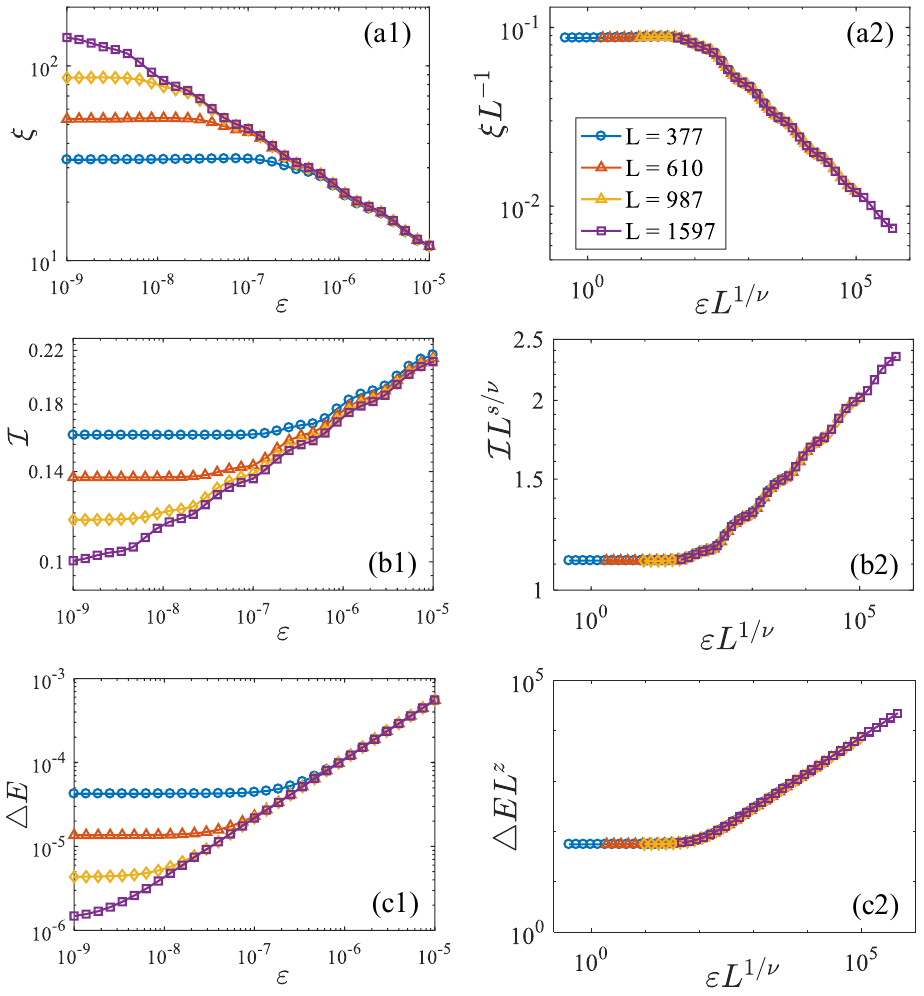}
	\caption{Critical properties in the AAS model with fixed $\delta L^{1/\nu_{\delta}}=1$. Curves of $\xi$ versus $\varepsilon$ before (a1) and after (a2) rescaling for different system sizes $L$. Curves of $\mathcal{I} $ versus $\varepsilon$ before (b1) and after (b2) rescaling for various $L$. Curves of $\Delta E $ versus $\varepsilon$ before (c1) and after (c2) rescaling for various $L$. The results are averaged over 1000 choices of $\phi$.}
    \label{fig5}
\end{figure}

\begin{figure}[tb]
	\centering
	\includegraphics[width=0.46\textwidth]{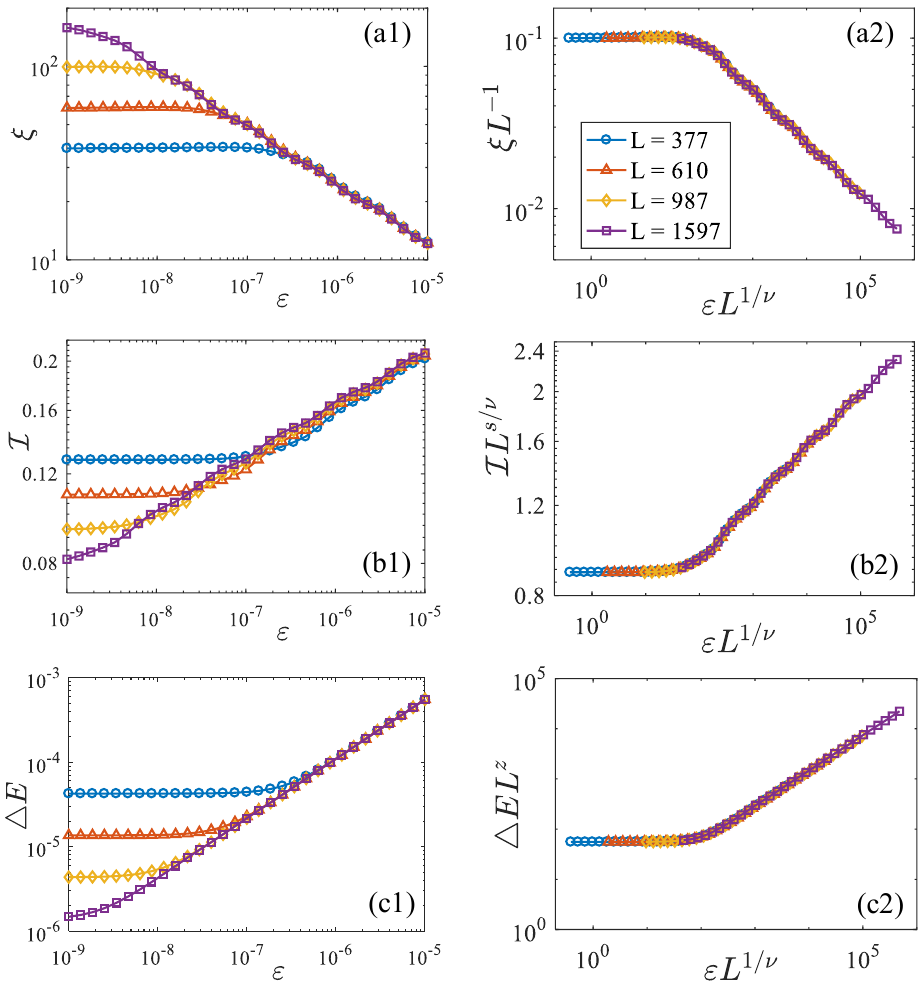}
	\caption{Critical properties in the AAS model with fixed $\delta L^{1/\nu_{\delta}}=-1$. Curves of $\xi$ versus $\varepsilon$ before (a1) and after (a2) rescaling for different system sizes $L$. Curves of $\mathcal{I} $ versus $\varepsilon$ before (b1) and after (b2) rescaling for various $L$. Curves of $\Delta E $ versus $\varepsilon$ before (c1) and after (c2) rescaling for various $L$. The results are averaged over 1000 choices of $\phi$.}
    \label{fig6}
\end{figure}

\subsection{\label{B}Aubry-Andr\'{e}-Stark criticality}

We now investigate the critical properties in the AAS model through analyzing the effect of the Stark field on the AA critical point. We first examine the physical quantities $P=\{\xi,\mathcal{I},\Delta E\}$ versus $g=\varepsilon$ at the critical point $\delta=0$. At this point, the finite-size scaling form of each physical quantity for different scales is obtained as
\begin{align}
\label{Eq:xiscaling4} &\xi=Lf_7(\varepsilon L^{1/\nu}),\\
\label{Eq:iprscaling5}&{\mathcal{I} }=L^{-s/\nu} f_8(\varepsilon L^{1/\nu}), \\
\label{Eq:gapscaling5}&\Delta E=L^{-z} f_9(\varepsilon L^{1/\nu}),
\end{align}
where $\{\nu,s,z\}$ denote the critical exponents for the AAS model and can be numerically determined from the data collapse.

The numerical results of localization length $\xi$ versus the Stark field strength $\varepsilon$ for different system sizes are shown in Fig.~\ref{fig4}(a1). By rescaling $\xi$ and $\varepsilon$ as $\xi L^{-1}$ and $\varepsilon L^{1/\nu}$ according to Eq.~(\ref{Eq:xiscaling4}), we find that all curves collapse into one curve when $\nu=0.3$, as shown in Fig.~\ref{fig4}(a2). Apparently, $\nu=0.3$ appears to be a new critical exponent in the AAS model, which is different from both $\nu_{\delta}=1$ for the pure AA model and $\nu_{\varepsilon}=0.33$ for the pure Stark model. This indicates that the Stark field contributes a new relevant direction at the AA critical point. Additionally, $\nu<\nu_{\delta}$ indicates that the Stark field is less relevant than the quasiperiodic potential. An explanation is that the Stark potential exhibits short-range correlation, while the quasiperiodic potential displays long-range correlation.

Figure \ref{fig4}(b1) shows the IPR $\mathcal{I}$ as a function of $\varepsilon$ for various system sizes. According to Eq.~(\ref{Eq:iprscaling5}), we rescale $\mathcal{I}$ and $\varepsilon$ as $\mathcal{I} L^{s/\nu}$ and $\varepsilon L^{1/\nu}$ in Fig.~\ref{fig4}(b2), which suggests that the best collapse of the curves using $s=0.098$ for the AAS model. This critical exponent is again different both from $s_{\delta}=0.333$ for the pure AA model and $s_{\varepsilon}=0.33$ for the pure Stark model. However, the ratio $s/\nu\approx0.33\approx s_{\delta}/\nu_{\delta}$ indicates that at the critical point, the scaling of the $\mathcal{I}$ with the system size given by Eq. (\ref{Eq:iprscaling1}) is the same for the pure AA model and the AAS model. Finally, the energy gap $\Delta E$ versus $\varepsilon$ for different system sizes is shown in Fig. \ref{fig4}(c1), and the rescaled curves are plotted in Fig. \ref{fig4}(c2). We determine the scaling exponent $z$ from the data collapse as $z=2.374=z_{\delta}$, which is also the same as that in the pure AA model.

We proceed to perform the scaling analysis in the critical region A with $\delta\neq 0$, as shown in Fig. \ref{fig1}(a). In the region A, the criticality of the AAS model is dependent on both $\delta$ and $\varepsilon$, such that the previous scaling forms for $\delta=0$ should be generalized. Concretely, we illustrate that the scaling behaviors of the characteristic quantities $P=\{\xi, \mathcal{I}, \Delta E\}$ introduced in Sec.~\ref{sec2} can be described by the scaling forms with $\delta$ and $\varepsilon$ as the scaling variables. In the critical region, we obtain the general finite-size scaling form of each quantity as
\begin{align}
	\label{Eq:xiscaling5} &\xi=L f_{10}(\delta L^{1/\nu_\delta},\varepsilon L^{1/\nu}),\\
	\label{Eq:iprscaling6}&\mathcal{I} =L^{-s/\nu} f_{11}(\delta L^{1/\nu_\delta},\varepsilon L^{1/\nu}), \\
	\label{Eq:gapscaling6}&\Delta E=L^{-z} f_{12}(\delta L^{1/\nu_\delta},\varepsilon L^{1/\nu}).
\end{align}
Based on the above general scaling forms, we can derive the same critical exponent $z=z_{\delta}$ and ratio $s/\nu=s_{\delta}/\nu_{\delta}$ for the AAS model and the pure AA model, which are only numerical results in previous sections. To do this, we first consider $\mathcal{I}$ in Eq. (\ref{Eq:iprscaling6}) for $\varepsilon=0$ and $L\rightarrow\infty$, which yields the scaling form $\mathcal{I} \propto \delta^{s \nu_{\delta}/\nu}$. Comparing with Eq.~(\ref{Eq:iprscaling2}), one obtains $s/\nu =s_{\delta}/\nu_{\delta}$. We then consider $\Delta E$ in Eq. (\ref{Eq:gapscaling6}) for $\varepsilon=0$ and $L\rightarrow\infty$, and obtain the scaling form $\Delta E \propto \delta^{\nu_{\delta}z}$. Comparing with $\Delta E \propto \delta^{\nu_{\delta}z_{\delta}}$ for the pure AA critical point, one has $z=z_{\delta}$ for the AAS model. To further validate the scaling forms in Eqs.~(\ref{Eq:xiscaling5},\ref{Eq:iprscaling6},\ref{Eq:gapscaling6}), we numerically calculate the scaling properties of $P=\{\xi, \mathcal{I}, \Delta E\}$ in the AA critical region for fixed $\delta L^{1/\nu_{\delta}}$=1 ($\delta >0$) in Fig.~\ref{fig5} and $\delta L^{1/\nu_{\delta}}=-1$ ($\delta <0$) in Fig.~\ref{fig6}, respectively. In both cases, we find the perfect collapse of rescaled curves according to Eqs.~(\ref{Eq:xiscaling5},\ref{Eq:iprscaling6},\ref{Eq:gapscaling6}), when we choose the same critical exponents $\{\nu,s,z\}=\{0.3,0.098,2.374\}$ with those for $\delta=0$ in Fig. \ref{fig4}.

\begin{figure}[tb]
	\centering
	\includegraphics[width=0.46\textwidth]{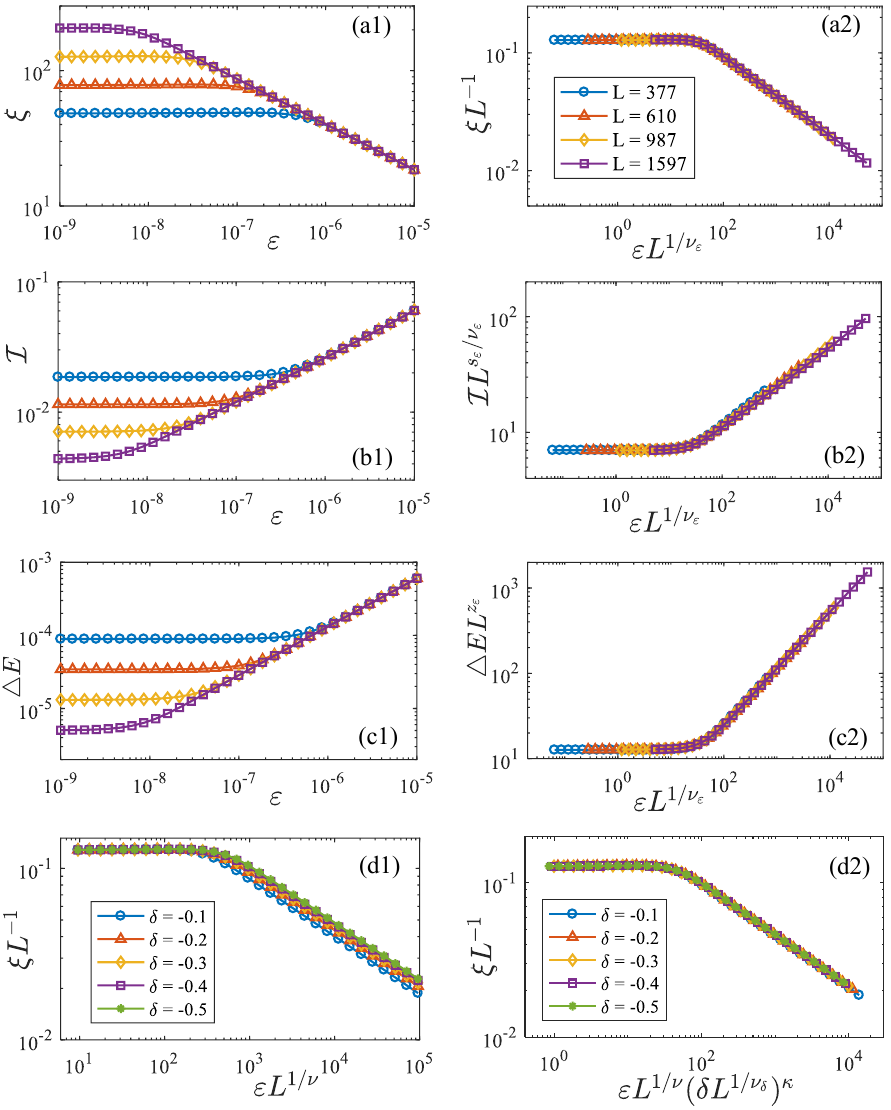}
	\caption{Critical properties in the AAS model with fixed $\delta =-0.1$. Curves of $\xi$ versus $\varepsilon$ before (a1) and after (a2) rescaling for different system sizes $L$. Curves of $\mathcal{I} $ versus $\varepsilon$ before (b1) and after (b2) rescaling for various $L$. The curves of $\Delta E $ versus $\varepsilon$ before (c1) and after (c2) rescaling for various $L$. (d1) Curves of $\xi L^{-1}$ versus $\varepsilon L^{1/v}$ for different values of $\delta$ with fixed $L=987$, and (d2) curves of $\xi L^{-1}$ versus $\varepsilon L^{1/v}(\delta L^{1/\nu_{\delta}})^{\kappa}$. The results are averaged over 1000 choices of $\phi$.}
    \label{fig7}
\end{figure}

Moreover, when $\delta<0$, there is an overlapping region between the critical region A for the AAS model and the critical region B for the pure Stark model, as shown in Fig. \ref{fig1}(a). As a result, one should impose a constraint on the scaling functions, which gives a hybrid scaling form \cite{S.Yin2022a}. To demonstrate this point, we fix $\delta=-0.1$ that is far away from the AA critical point. In Fig.~\ref{fig7}(a1), we compute $\xi$ versus $\varepsilon$ for different system size. After rescaling $\xi$ and $\varepsilon$ as $\xi L^{-1}$ and $\varepsilon L^{1/\nu}$, we find that the rescaled curves collapse well by setting the same critical exponent $\nu_{\varepsilon}=0.33$ as that for the Stark localization transition, as shown in Fig.~\ref{fig7}(a2). Similarly, we numerical calculate $\mathcal{I}$ and $\Delta E$ versus $\varepsilon$ in Figs.~\ref{fig7}(b1) and \ref{fig7}(c1), respectively. Apparently, the rescaled curves in Figs.~\ref{fig7}(b2) and \ref{fig7}(b2) collapse well if we choose the same critical exponents $s_{\varepsilon}=0.33$ and $z_{\varepsilon}=2$ as those for the pure Stark model. This indicates that the scaling forms in Eqs.~(\ref{Eq:xiscaling3},\ref{Eq:iprscaling4},\ref{Eq:gapscaling4}) are still valid in the overlapping critical region. Thus, in this region, the scaling behaviors can be simultaneously described by the scaling forms of the AAS model and the pure Stark model.
In particular, the localization length $\xi$ satisfies both the scaling forms in Eq.~(\ref{Eq:xiscaling3}) and Eq.~(\ref{Eq:xiscaling5}). This yields a hybrid scaling form of $\xi$ as
\begin{equation}
    \label{Eq:xiscaling7}
    \xi = L f_{13} (\varepsilon L^{\frac{1}{\nu}} (\delta L^{\frac{1}{\nu_{\delta}}})^{\kappa})
 \end{equation}
where $\kappa=\nu_{\delta}/\nu_{\varepsilon}-\nu_{\delta}/\nu=-0.303$. We show the numerical results of $\xi L^{-1}$ as a function of $\varepsilon L^{1/\nu}$ for various $\delta$ and fixed $L=987$ in Fig. \ref{fig7}(d1). By using the collapse plot of $\xi L^{-1}$ versus the hybrid quantity $\varepsilon L^{{1}/{\nu}} (\delta L^{{1}/{\nu_{\delta}}})^{\kappa}$ in Fig. \ref{fig7}(d2), we find all curves collapse onto one curve by setting $\kappa = -0.303$. This confirms the hybrid scaling form of the localization length in Eq.~(\ref{Eq:xiscaling7}).

\section{\label{sec4}KZS of driven dynamics}

We turn to investigate the KZS of the driven dynamics in the AAS model, which is closely related to the quantum criticality of phase transitions \cite{dziarmaga2010dynamics,Polkovnikov2011}. We consider the system is initially in the localization state and then is driven to pass through the critical point by linearly varying the distance $g$ in time $t$ with the speed $R$. The time evolution of $g$ is given by
\begin{equation}
 \label{Eq:kzs}
   g(t)=g_{0}-Rt,
\end{equation}
where $g$ can be $\delta$ or $\varepsilon$ depending on the model considered, and $g_{0}>0$ represents the initial distance from the critical point at $t=0$. According to the KZS, when $|g|>R^{1/r\nu}$ with the scaling exponent $r=z+1/\nu$, the system has enough time to adjust the change of the Hamiltonian to preserve adiabatic; while when $|g|<R^{1/r\nu}$, the change rate of the system itself is less than that of the external parameter, which implies the system entering the impulse region. Below, we choose the system size $L=1597$, which is sufficiently large to ignore the finite-size effect in real-time simulations. We set the initial state as the ground state of the system with fixed $\phi=0.2$ and $g_0=1.5$.

\begin{figure}[tb]
	\centering
	\includegraphics[width=0.46\textwidth]{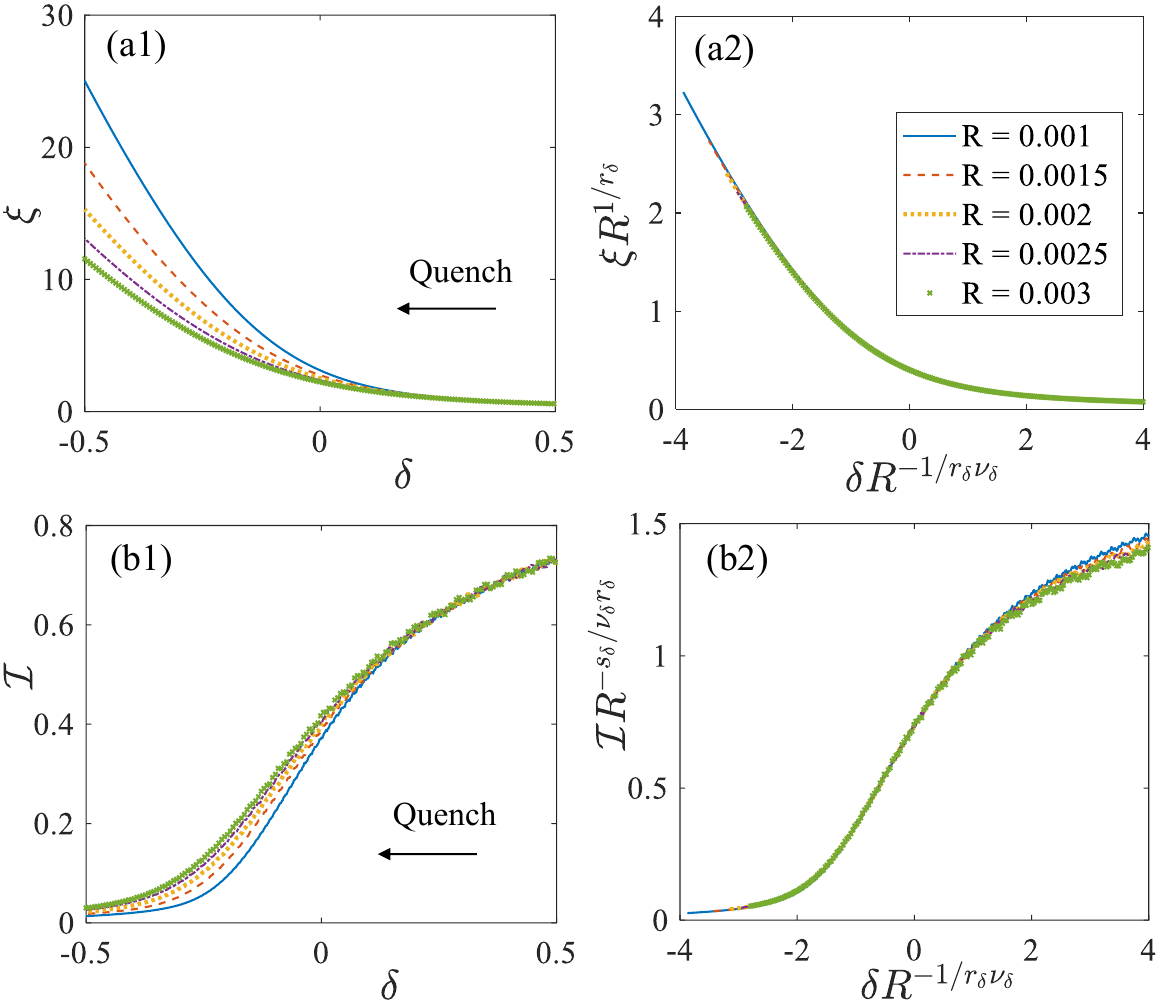}
	\caption{KZS of driven dynamics in the AA model with $\varepsilon=0$. Curves of $\xi$ versus $\delta$ before (a1) and after (a2) rescaling for different driving rates $R$. Curves of $\mathcal{I} $ versus $\delta$ before (b1) and after (b2) rescaling for different $R$. The black arrows denote the quench direction.}
	\label{fig8}
\end{figure}

We first study the KZS in the pure AA model with $\varepsilon=0$, $g=\delta$, and critical exponents $\{\nu_{\delta},z_{\delta}\}=\{1,2.374\}$. In this limit and when the system size is sufficiently large, the KZS form of the localization length $\xi$ near the critical point is given by
\begin{eqnarray}
    \label{Eq:dynamicScalingXiR}
    \xi(\delta,R) &=& R^{-1/r_{\delta}}f_{14}(\delta R^{-1/r_{\delta}\nu_{\delta}}),
   \end{eqnarray}
where $r_{\delta}=z_{\delta}+1/\nu_{\delta}$ is denoted for the pure AA model. The numerical result of $\xi$ versus $\delta$ for different driving rates $R$ is shown in Fig.~\ref{fig8}(a1). We find that when $\delta > R^{1/r_{\delta}\nu_{\delta}}$, $\xi$ is independent of $R$, such that the curves under different $R$ coincide. This implies the adiabatic evolution of the system at this time. When $\delta < R^{1/r_{\delta}\nu_{\delta}}$, the system enters the impulse region and the curves under different $R$ are separated from each other. By rescaling $\xi$ and $\delta$ as $\xi R^{1/r_{\delta}}$ and $ \delta R^{-1/r_{\delta}\nu_{\delta}}$ in Fig.~\ref{fig8}(a2), we find that the curves of different $R$ collapse well near the critical point, which confirms the KZS form given by Eq. (\ref{Eq:dynamicScalingXiR}). Similarly, the IPR $\mathcal{I} $ satisfies the KZS form
\begin{equation}
\label{Eq:dynamicScalingipr}
    {\mathcal{I}}(\delta,R)=R^{s_{\delta}/r_{\delta} \nu_{\delta}}f_{15}(\delta R^{-1/r_{\delta}\nu_{\delta}}).
\end{equation}
As shown in Fig.~\ref{fig8}(b1), when $\delta > R^{1/r_{\delta}\nu_{\delta}}$ ($\delta < R^{1/r_{\delta}\nu_{\delta}}$ near the critical point) for the adiabatic (impulse) region, the curves of $\mathcal{I}$ for different driving rates $R$ coincide (are separated from each other).
By rescaling $ \mathcal{I} $ and $\delta$ as $ \mathcal{I}  R^{-s_{\delta} /r_{\delta}\nu_{\delta} }$ and $ \delta R^{-1/r_{\delta}\nu_{\delta}}$ in Fig.~\ref{fig8}(b2), we find that the curves collapse according to Eq. (\ref{Eq:dynamicScalingipr}).

\begin{figure}[tb]
	\centering
	\includegraphics[width=0.46\textwidth]{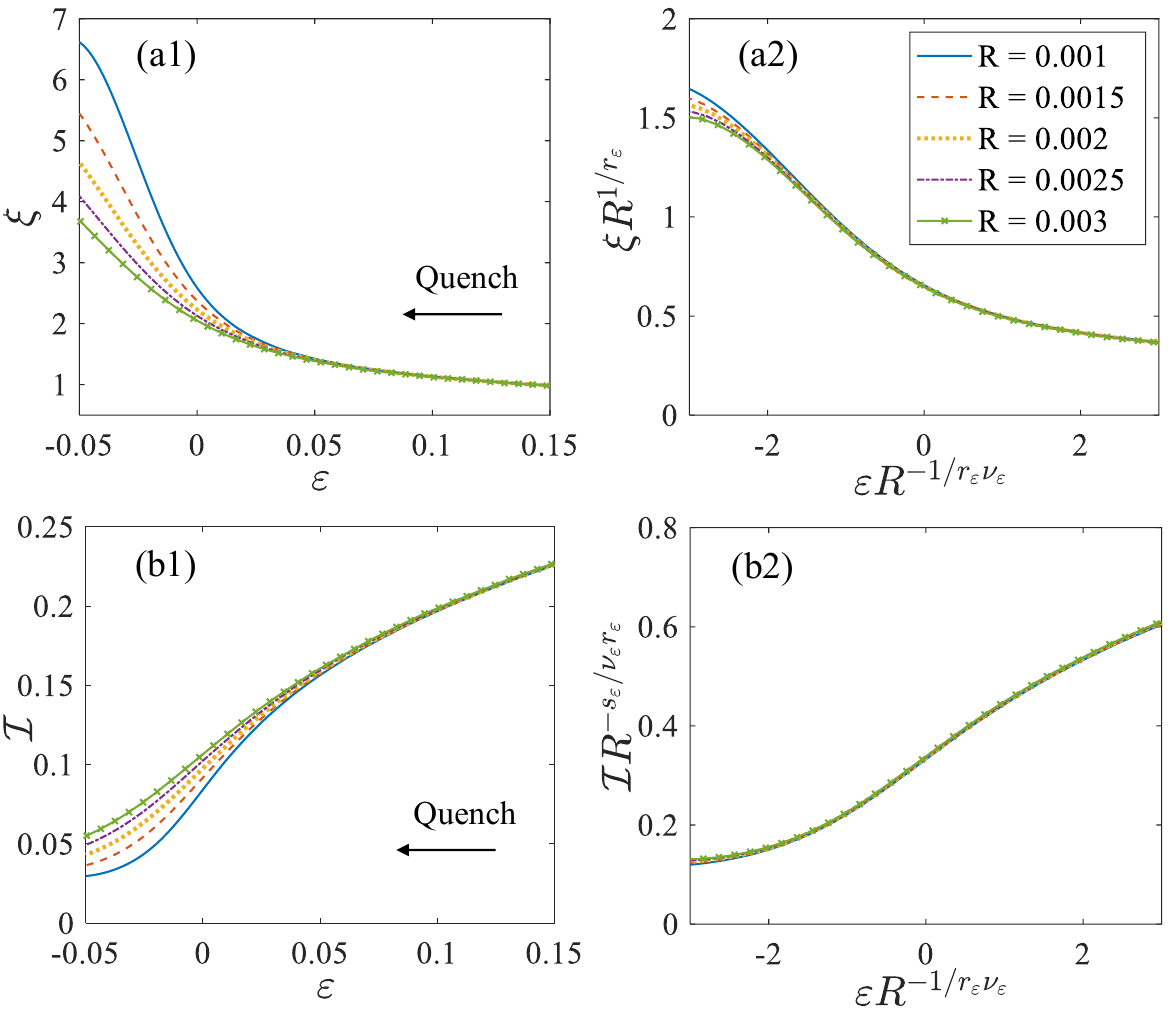}
	\caption{KZS of driven dynamics in the Stark model with $\delta =-2J$. Curves of $\xi$ versus $\varepsilon$ before (a1) and after (a2) rescaling for different driving rates $R$. Curves of $\mathcal{I} $ versus $\varepsilon$ before (b1) and after (b2) rescaling for different $R$. The black arrows denote the quench direction.}
    \label{fig9}
\end{figure}

We then study the KZS in the pure Stark model with $\delta=-2J$ and $g=\varepsilon$. In this case, the KZS form of $\xi$ and $\mathcal{I}$ are given by
\begin{align}
	\label{Eq:dynamicScalingXiR1} & \xi(\varepsilon,R) = R^{-1/r_{\varepsilon}}f_{16}(\varepsilon R^{-1/r_{\varepsilon}\nu_{\varepsilon}}),\\
    \label{Eq:dynamicScalingipr1} &\mathcal{I}(\varepsilon,R)=R^{s_{\varepsilon}/r_{\varepsilon} \nu_{\varepsilon}}f_{17}(\varepsilon R^{-1/r_{\varepsilon}\nu_{\varepsilon}}),
\end{align}
respectively. The numerical results of $\xi$ and $\mathcal{I}$ versus $\varepsilon$ for different driving rate $R$ are shown in Fig.~\ref{fig9}(a1) and \ref{fig9}(b1), respectively. The initial evolutions of $\xi$ and $\mathcal{I}$ are independent of $R$ when $\varepsilon > R^{1/r_{\varepsilon}\nu_{\varepsilon}}$. Near the critical point with $\varepsilon < R^{1/r_{\varepsilon}\nu_{\varepsilon}}$, the system enters the impulse region, and the curves of $\xi$ and $\mathcal{I}$ for different driving rate $R$ are separated from each other. By using the collapse
plot of $\xi R^{1/r_{\varepsilon}}$ ($\mathcal{I}  R^{-s_{\varepsilon} /r_{\varepsilon}\nu_{\varepsilon} }$) and $\varepsilon R^{-1/r_{\varepsilon}\nu_{\varepsilon}}$ ($\varepsilon R^{-1/r_{\varepsilon}\nu_{\varepsilon}}$) in Fig.~\ref{fig9}(a2) [Fig.~\ref{fig9}(b2)], we confirm the KZS of $\xi$ ($\mathcal{I}$) in Eq. (\ref{Eq:dynamicScalingXiR1}) [Eq. (\ref{Eq:dynamicScalingipr1})] for the pure Stark model with the critical exponents $\nu_{\varepsilon}=0.33$ and $z_{\varepsilon}=2$.

 \begin{figure}[tb]
	\centering
	\includegraphics[width=0.46\textwidth]{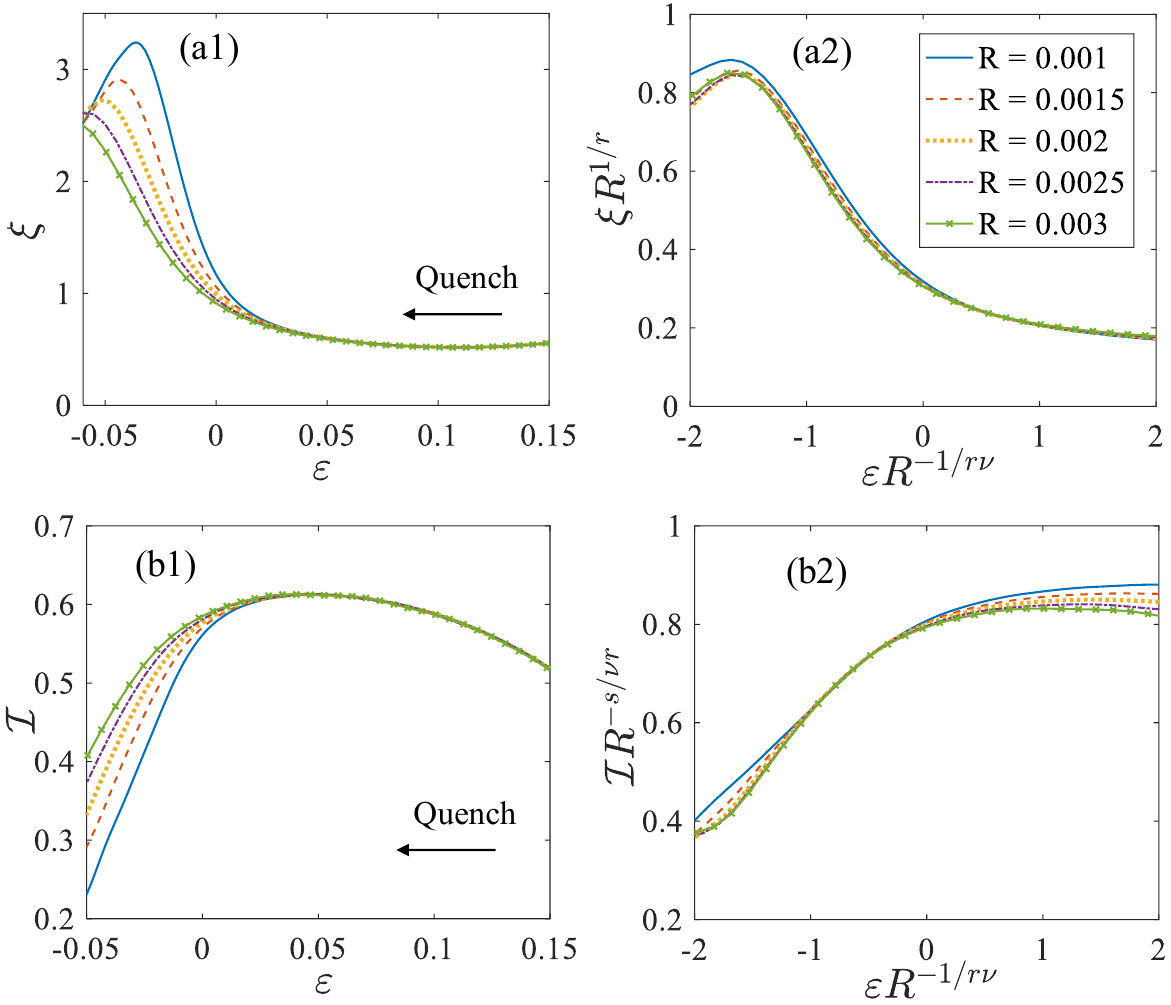}
\caption{KZS of driven dynamics in the AAS model with fixed $\delta=0$. Curves of $\xi$ versus $\varepsilon$ before (a1) and after (a2) rescaling for different driving rates $R$. Curves of $\mathcal{I} $ versus $\varepsilon$ before (b1) and after (b2) rescaling for different $R$. The black arrows denote the quench direction.}
\label{fig10}
\end{figure}

\begin{figure}[tb]
	\centering
	\includegraphics[width=0.46\textwidth]{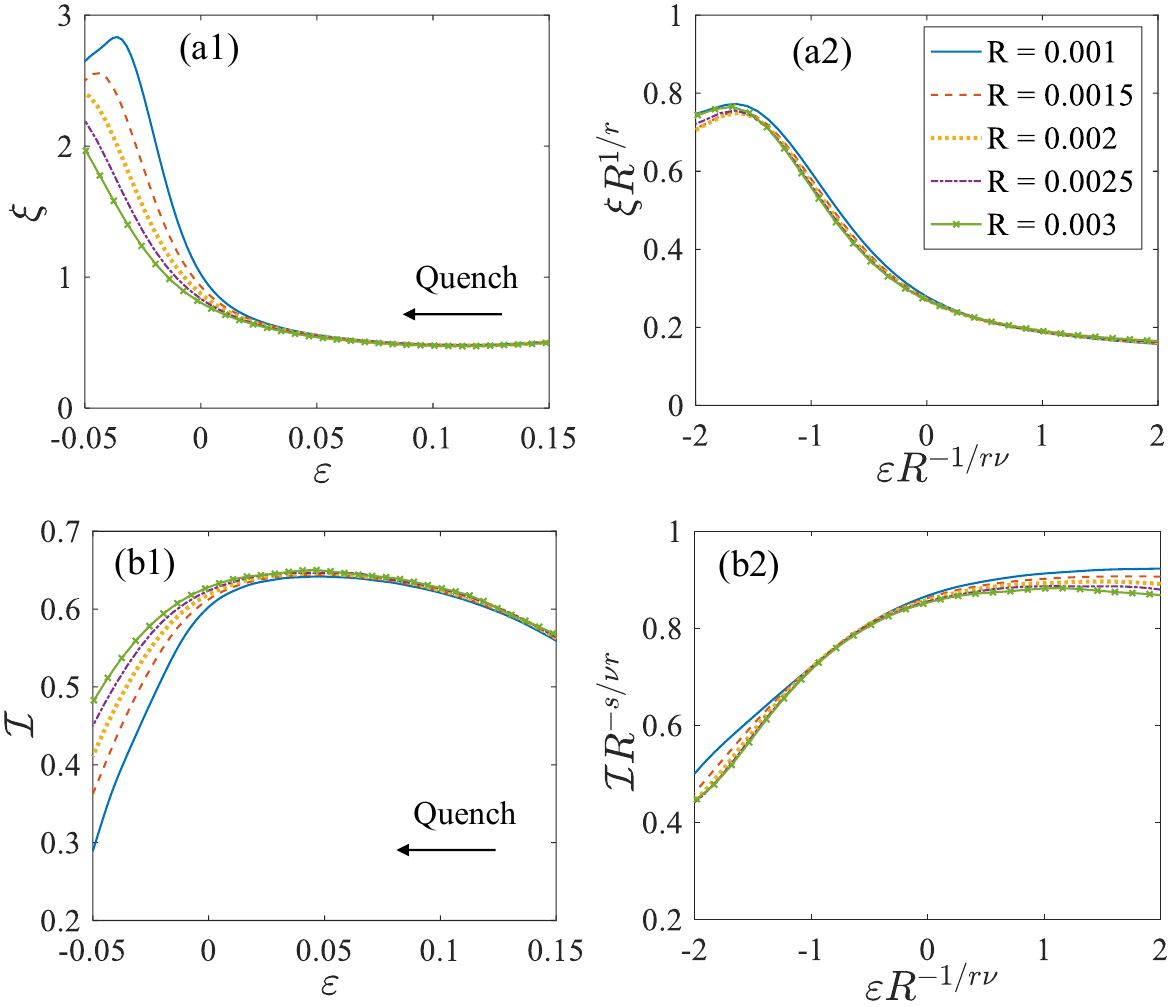}
\caption{KZS of driven dynamics in the AAS model with fixed $\delta R^{-1/r\nu_{\delta}}=0.3$. Curves of $\xi$ versus $\varepsilon$ before (a1) and after (a2) rescaling for different driving rates $R$. Curves of $\mathcal{I} $ versus $\varepsilon$ before (b1) and after (b2) rescaling for different $R$. The black arrows denote the quench direction.}
\label{fig11}
\end{figure}

Finally, we explore the KZS in the general AAS model. Apparently, there are two adjustable variables $\delta$ and $\varepsilon$ in the model. In this case, the full KZS form of the physical quantities $P=\{\xi, \mathcal{I}\}$ can be written as
\begin{align}
    \label{Eq:dynamicScalingXiR2} &\xi(\varepsilon,\delta,R) = R^{-1/r}f_{18}(\varepsilon R^{-1/r\nu},\delta R^{-1/r\nu_\delta}),\\
    \label{Eq:dynamicScalingipr2} &\mathcal{I}(\varepsilon,\delta,R) = R^{s /r \nu}f_{19}(\varepsilon R^{-1/r\nu},\delta R^{-1/r\nu_\delta}),
\end{align}
where $r=z+1/\nu$ with $\nu=0.3$ and $z=2.374$ for the AAS model. When $\delta = 0$, Eq.~(\ref{Eq:dynamicScalingXiR2}) and Eq.~(\ref{Eq:dynamicScalingipr2}) return to the simplified forms
\begin{align}
    \label{Eq:dynamicScalingXiR3} &\xi(\varepsilon,R)= R^{-1/r}f_{20}(\varepsilon R^{-1/r\nu}),\\
    \label{Eq:dynamicScalingipr3} &\mathcal{I}(\varepsilon,R) = R^{s /r \nu}f_{21}(\varepsilon R^{-1/r\nu}).
\end{align}
The numerical results of $\xi$ and $\mathcal{I}$ versus $\varepsilon$ for various driving rates $R$ are shown in Figs.~\ref{fig10}(a1) and \ref{fig10}(b1), respectively. One can observe that the curves under different $R$ separated from each near the critical point. After rescaling the curves with the critical exponents in AAS model, we find the rescaled curves collapses with each other near the critical point, as shown in Figs.~\ref{fig10}(a2) and \ref{fig10}(b2).
For $\delta\neq 0$, we can fix the value of $\delta R^{-1/r\nu_{\delta}}$ to verify the KZS forms of $\xi$ and $\mathcal{I}$, as shown in Fig. \ref{fig11}. Here we set $\delta R^{-1/r\nu_{\delta}} =0.3$ and numerically calculate the time evolutions of $\xi$ and $\mathcal{I}$ for various driving rate $R$, with results shown in Figs.~\ref{fig11}(a1) and \ref{fig11}(b1). After rescaling of the physical quantities according to Eq. (\ref{Eq:dynamicScalingXiR3}) and Eq. (\ref{Eq:dynamicScalingipr3}), one can see that the curves under different $R$ collapse together near the critical point in Figs.~\ref{fig11}(a2) and \ref{fig11}(b2). This demonstrates the KZS near the AAS critical point. We also numerically confirm the KZS for the AAS model with fixed $\delta R^{-1/r\nu_{\delta}}=-0.3$.

\section{\label{sec5}Conclusion}

In summary, we have systematically investigated the critical properties and the KZS in the AAS model. We have numerically calculate the localization length, the IPR, and the energy gap, and performed the scaling analysis to characterize quantum criticality of the localization transition. We have obtained the scaling forms of these characteristic physical quantities for the pure AA model, the pure Stark model, and the AAS model with new critical exponents that are different from the counterparts for both the former models. We have also revealed rich critical phenomena in the critical region spanned by the quasiperiodic and the Stark potentials, and obtained a hybrid scaling form in the overlapping critical regions of the Anderson and Stark localizations. Furthermore, we have explored the driven dynamics of the localization transitions in the AAS model. By linearly changing the strength of the Stark (or quasiperiodic) potential, we have calculated the evolution of the localization length and the IPR, and studied their dependence on the driving rate. We have found that the KZS describes well the driven dynamics from the ground state with the critical exponents obtained from the static scaling analysis, which can include two scaling variables when both the Stark and quasiperiodic potentials are relevant.

{\sl Note added}. Very recently, we noticed a preprint on quantum criticality in the AAS model \cite{sahoo2024stark}, where similar critical exponents were obtained. In our present work, we furthermore study the KZS of driven dynamics in the AAS model.

\begin{acknowledgments}
This work was supported by the National Natural Science Foundation of China (Grant No. 12174126), the Guangdong Basic and Applied Basic Research Foundation (Grant No. 2024B1515020018), and the Science and Technology Program of Guangzhou (Grant No. 2024A04J3004).
\end{acknowledgments}

\normalem
\bibliography{reference}

\end{document}